\documentclass[12pt]{article}

\usepackage{epsfig}
\usepackage{cite}

\title{Pion Production in Proton Collisions With Light Nuclei:
Implications for Atmospheric Neutrinos}

\author{Ralph Engel, T.K. Gaisser \& Todor Stanev\\
Bartol Research Institute\\
Newark, DE 19716 USA}

\date{November 16, 1999}

\begin{document}

\maketitle

\begin{abstract}
Differences among calculations of the atmospheric neutrino beam
can be traced
in large part to differences in the representation of pion
production by protons interacting with nuclei in the atmosphere.
In this paper we review the existing data with the goal of
determining the regions of phase space in which new measurements
could help to improve the input to the calculations.
\vspace*{-14cm}
\flushright{preprint BA-99-73}
\vspace*{13.5cm}
\end{abstract}


\section{Introduction}

\begin{figure}[!htb]
\centerline{\epsfig{figure=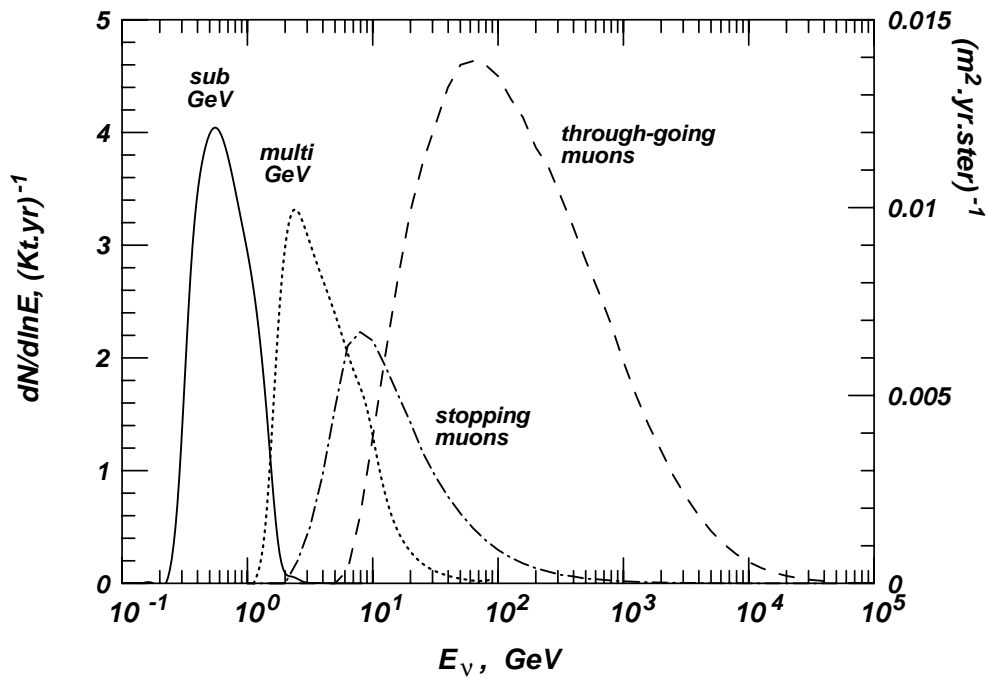,height=10cm}}
\caption{\label{fig1} 
 Distributions of neutrino energies that give rise to
 four classes of events.  Sub-GeV and multi-GeV refer to
 the two classes of contained events at Superkamiokande.
 The response for contained events is averaged over all
 angles, while  for stopping and throughgoing muons
it is only given for vertical upward going neutrino
 induced muons.
}
\end{figure}

The observed up-down asymmetry of the flux of muon-like events
induced by interactions of atmospheric neutrinos
has been interpreted as evidence for neutrino
oscillations~\cite{Fukuda98a}.
The case for neutrino oscillations is especially strong for
the multi-GeV events at Super-Kamiokande because the energies are
 high enough so that the charged leptons follow the neutrino direction
 closely and expose the oscillation effects as a function of the
 neutrino pathlength.
 On the other hand, comparisons to calculations of the atmospheric
 neutrino flux over as broad an energy range as possible are essential
 for making a detailed interpretation of the data and generally for
 giving confidence in the basic result.  In addition, calculations play
 a crucial role in comparison of measurements made at different locations
 and with different techniques.  In particular, the Soudan
 experiment~\cite{Allison97a}, and IMB~\cite{Becker-Szendy92} previously,
 detect neutrino interactions in a range similar to Super-Kamiokande
 (and Kamiokande~\cite{Kamiokande}),
 but the geomagnetic environments are quite different, which changes 
 the neutrino angular distribution expected in the absence of
 oscillations. Measurements of neutrino-induced upward muons,
 as also at MACRO~\cite{MACRO}, are also sensitive to oscillations,
 but in a different energy regime (see Fig.~\ref{fig1}).

There are several independent calculations of the flux of atmospheric
neutrinos~\cite{Agrawal96a,HKKM,Bugaev89a,Perkins,Battistoni,Waltham}.  In all
cases but one~\cite{Perkins}, the calculations start with the primary
cosmic-ray spectrum as measured (protons and alpha-particles give
the dominant contributions) and simulate the resulting cosmic-ray cascade
in the atmosphere, calculating the fluxes of $\nu_e$, $\bar{\nu}_e$,
$\nu_\mu$, $\bar{\nu}_\mu$, $\mu^+$, and $\mu^-$.  The calculated muon
fluxes can be compared with measurements of muons at high
altitude~\cite{MASS1,MASS2,HEAT,IMAX,CAPRICE} as a check of the calculation.
Perkins~\cite{Perkins} starts from the measured muon flux and uses
the kinematic relationship between muons and neutrinos to derive
the neutrino flux.  A limitation of the muon fluxes is that the
individual measurements generally have low statistics and depend
more sensitively than the neutrinos on the details of geomagnetic and
three-dimensional effects.

The most recent calculations~\cite{Battistoni,Waltham} constitute
a very significant technical advance in that they are three-dimensional.
In previous calculations secondary particles had been assumed to follow
the direction of the primary cosmic-ray that generated them.  A major
conclusion of Ref.~\cite{Battistoni} is that for practical purposes
the one-dimensional calculations give adequate results.  This is
because Fermi-momentum of the target nucleons of the neutrino interactions,
compounded by limited angular resolution, smear out angular features
visible in fluxes of $\sim$GeV neutrinos.   A corollary of this conclusion
is that simpler one-dimensional calculations can be used for
comparison of calculations in order to trace sources of
differences among the calculated neutrino fluxes.

 Some time ago the existing one-dimensional calculations were
 compared~\cite{Gaisser96a}.  The dominant sources of difference
were found to be the assumed primary spectrum and the representation
of pion production in collisions of protons with light nuclei.
Although differences in input were at the level of 20-30\% -- or
larger in some cases -- the neutrino fluxes of the two
calculations~\cite{HKKMold,Bartol}
that have been used extensively for input to the
detector simulations differ by much less.  
This is a consequence of compensating uncertainties together with
the fact that both calculations were constrained to fit the same
(ground-level) muon data.  Recently, several new measurements of
the primary spectrum~\cite{MASSp,CAPRICEp,BESS} confirm the
lower~\cite{Seo} of two older measurements~\cite{Webber,Seo}, leaving
differences in representation of pion production as the major
source of uncertainty for GeV neutrino fluxes.

A striking example is the difference between the calculations
of Battistoni {\it et~al.}~\cite{Battistoni} and the Bartol
fluxes~\cite{Agrawal96a}.  This comparison, which was made~\cite{TAUP}
using the one-dimensional version of Ref.~\cite{Battistoni} and the
same assumed primary spectrum as Ref.~\cite{Agrawal96a}, gives neutrino
fluxes 20-30\% lower than Ref.~\cite{Agrawal96a} for $E_\nu<10$~GeV.
A detailed comparison with the
new calculation of Ref.~\cite{Battistoni} is currently in
progress~\cite{Battistoni2}.


\section{Inclusive pion production around 20 GeV}

  To see what range of interaction energies is most important
for the various classes of events shown in Fig.~\ref{fig1}, one
needs to look at the distributions of primary energies that
produce the events.  Roughly speaking, for the steep
cosmic ray spectrum the most relevant primary energies are
an order of magnitude higher than the neutrino energy of interest.
For the upward neutrino-induced muons, for example, the important range of
interaction energies extends up to several TeV.  In this energy
range, the uncertainties in the primary spectrum are still relatively
large, and there are also significant uncertainties from the
amount and momentum distribution of kaon production~\cite{GLS},
which will be addressed in~\cite{Battistoni2}.
Here we concentrate on the lower energy events (e.g. the sub-GeV
events at Super-Kamiokande) for which most of the contribution
comes from interactions of primary cosmic-ray nucleons with
energies between 10 and 100 GeV.  A similar range of energies
is responsible for the Soudan events and for contained events
in IMB and Kamiokande.

\begin{figure}[!htb]
\centerline{\epsfig{figure=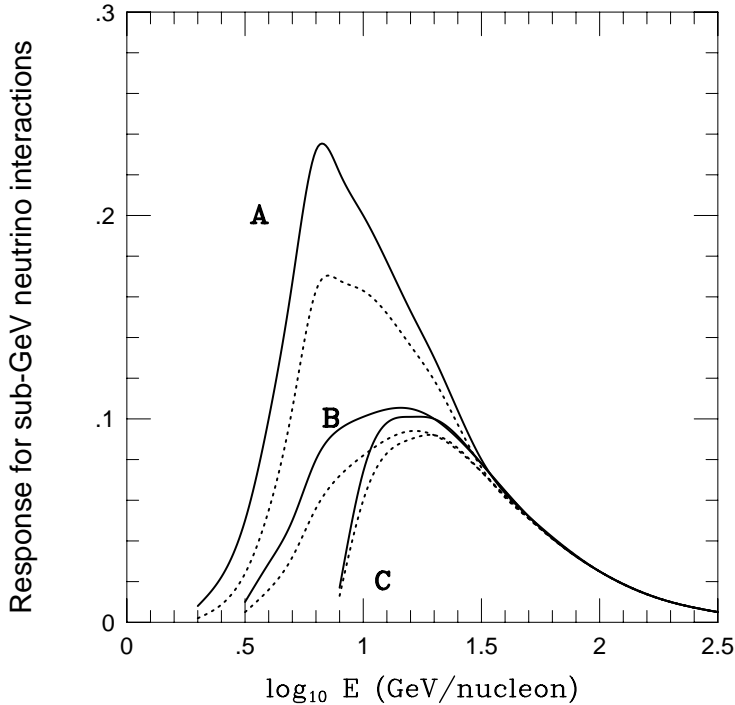,height=11cm}}
\caption{\label{response}
 Response for sub-GeV muon-like events
 in Super-K to the energy of the primary cosmic
 ray nucleons.
 A: no geomagnetic cutoff; 
 B: events from lower hemisphere (upward going leptons);
 C: events from upper hemisphere (downward going leptons).
 Each pair of curves shows the range of the signal between
 minimum (solid) and maximum (dotted) of solar activity.
}
\end{figure}
We show the distribution of primary energies
that gives rise to the sub-GeV events at Super-Kamiokande
in Fig.~\ref{response}~\cite{nu98}.  We show separately the response for
downward (C) and upward (B) events, as well as the response
that would apply if there were no geomagnetic field (A).
It is interesting to note that the geomagnetic cutoff leads to an
observable site-dependence of the ratio of downward to upward moving
events.  In the absence of oscillations, due to the higher local
geomagnetic cutoff at Kamioka in contrast to the very low
vertical cutoff in the northern U.S., the ratio of downward to
upward going leptons is respectively less and greater than one.

\begin{figure}[!htb]
\centerline{\epsfig{figure=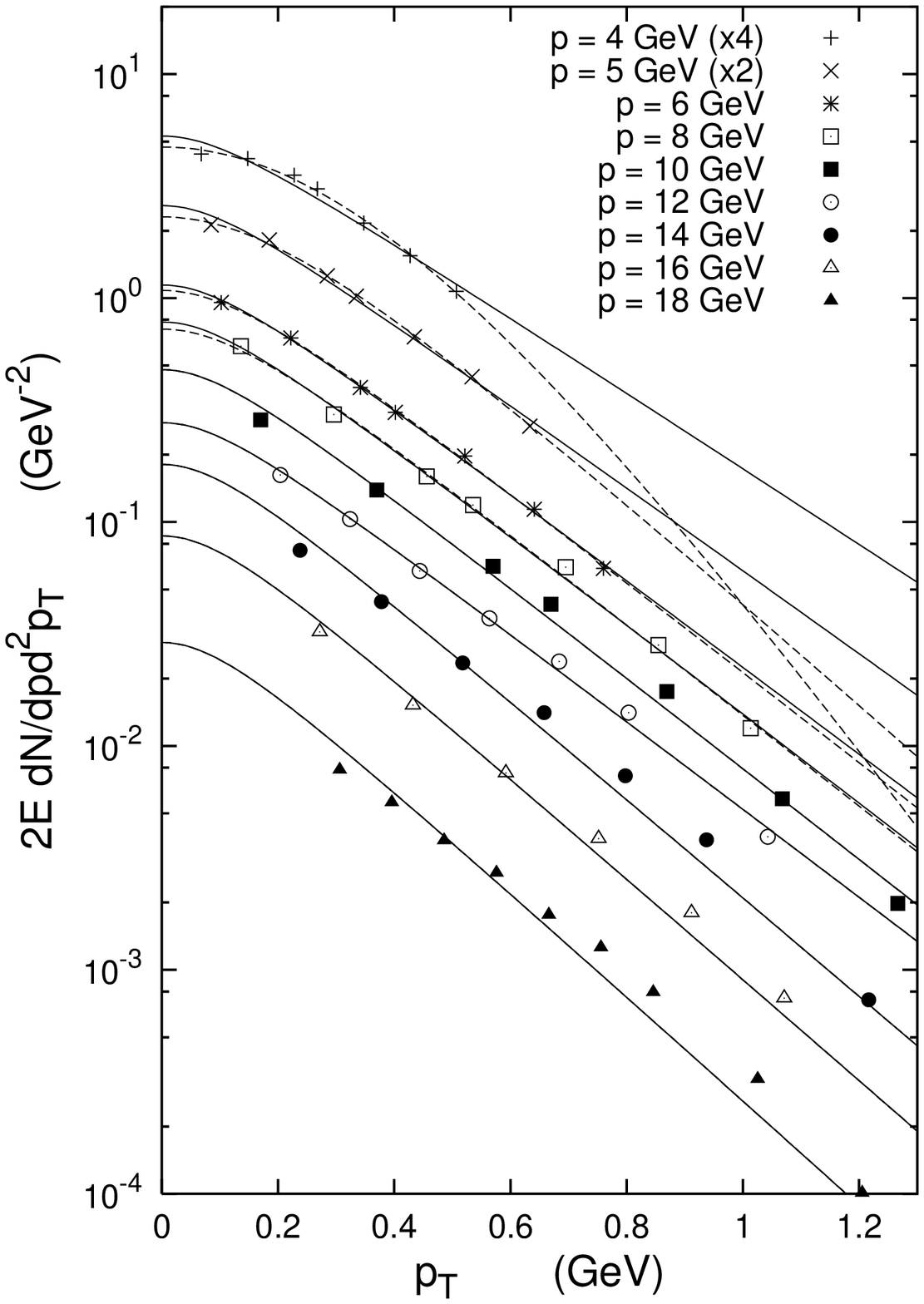,height=14cm}}
\caption{\label{Eichten-fit-pip-fig} Data on inclusive $\pi^+$
production in proton-beryllium collisions at 24 GeV \cite{Eichten72}. 
The solid lines
correspond to fits using the pion mass in (\ref{trans-mass-eq}) and the
dashed curves show the result if the mass is treated as free parameter.
}
\end{figure}
One of the most extensive data sets in the energy range responsible for
sub-GeV events is
the work of Eichten {\it et~al.}~\cite{Eichten72}.  Figure 3 shows
the data from this experiment for production of $\pi^+$ from interactions
of 24 GeV/c protons on beryllium.  The solid lines are fits of the form
\begin{equation}
2 E{dN\over dp d^2p_T}\;=\;a(p)\times
\exp\left\{-b(p)\; m_T\right\},
\label{Eqn1}
\end{equation}
where the transverse mass is defined as 
\begin{equation}
m_T=\sqrt{p_T^2+m_\pi^2},
\label{trans-mass-eq}
\end{equation}
and $a(p)$, $b(p)$ are free parameters at each value of pion
momentum $p$.  This form gives good fits for all 
momenta except the lowest two values.  The dashed lines show fits
in which the pion mass entering Eq.~(\ref{trans-mass-eq}) 
is also treated as a free parameter.
For negative pions, the form (\ref{Eqn1}) gives good fits for all
measured momenta.

\begin{figure}[!htb]
\centerline{\epsfig{figure=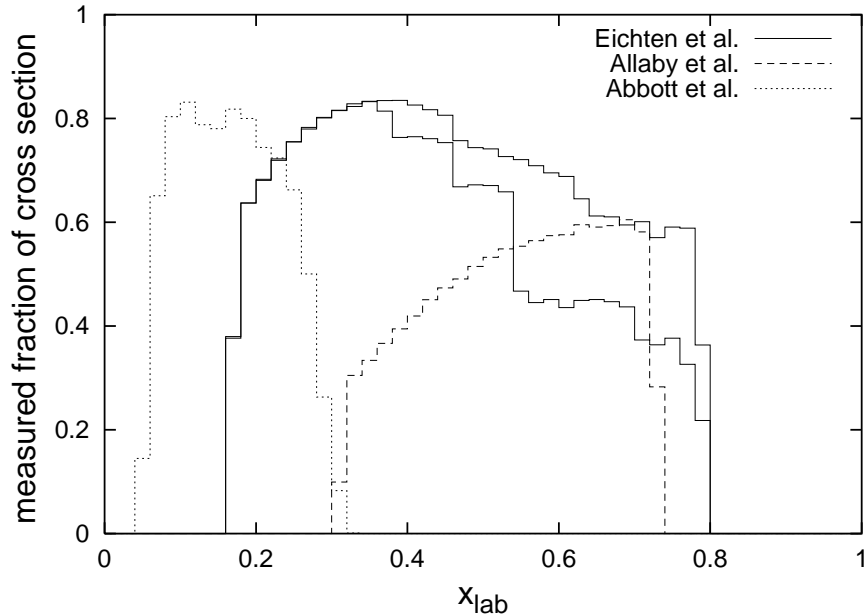,width=12cm}}
\caption{\label{all-ps-fig} Fraction of pion production cross
section covered by the data of
Eichten {\it et~al.} (p-Be at 24 GeV \protect\cite{Eichten72}),
Allaby {\it et~al.} (p-Be at 19.2 GeV \protect\cite{Allaby70}),
and Abbott {\it et~al.} (p-Be at 14.6 GeV \protect\cite{Abbott92}).
For Eichten {\it et~al.} the upper and lower curves show the coverage
for $\pi^+$ and $\pi^-$ respectively. The other experiments have
acceptances which are essentially identical for negatively and
positively charged pions.}
\end{figure}

 What is important for the overall normalization of the 
 atmospheric  neutrino flux is the total yield of pions
 integrated over all phase space.  This involves both
 interpolation and extrapolation. Figure~\ref{all-ps-fig}
 illustrates the fraction of the pion production cross section
 that requires interpolation only for the data of Ref.~\cite{Eichten72}.
The curve represents the fraction of secondary pions 
falling into the acceptance range of the experiment as function of
$x_{\rm lab} = p/p_{\rm beam}$, using the predictions of the 
TARGET Monte Carlo \cite{Gaisser83a}.

Thus, for example, for $\pi^+$ at $p = 6$~GeV ($x_{\rm lab}\approx 0.25$)
approximately 75\% of the integral over transverse momentum does not
require extrapolation into unmeasured regions of transverse momentum.
Most important is the fact that in this experiment there are no
data at all for $p<4$~GeV. In this region,
therefore, the representation of pion production depends on
how well the model extrapolates in both longitudinal and transverse momentum.

Other experiments~\cite{Lundy65,Allaby70,Abbott92}
cover various regions of phase space at various
nearby beam energies (see Fig.~\ref{all-ps-fig}).
The Lundy {\it et~al.} data set \cite{Lundy65} (p-Be at 12.5 GeV)
is not shown since it
corresponds to pion production spectra which are not consistent with the
other measurements.

 Only the data of Abbott {\it et~al.} \cite{Abbott92} 
 reach into the $x_{\rm lab}$ region of about 0.1 to 0.2,
 which is an important contributor to the atmospheric 
 lepton fluxes. Since these measurements are published as
 function of particle rapidity $y = \log[(E+p_L)/(E-p_L)]$
 and transverse mass $m_T$, the conversion of the data into
 $dN/dx_{\rm lab}$ requires a two-dimensional inter- and extrapolation
 resulting in substantial uncertainties.

\begin{figure}[!htb]
\centerline{\epsfig{figure=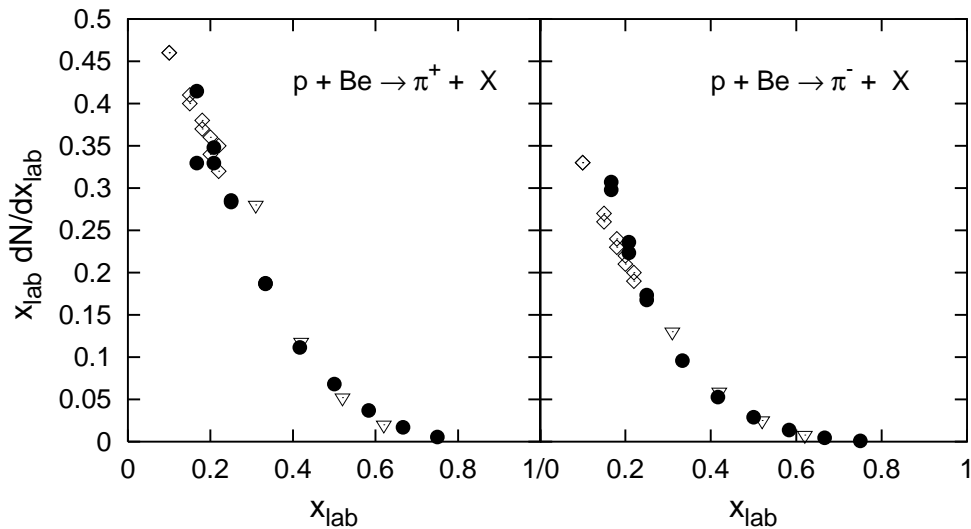,width=13cm}}
\caption{\label{xdndx-all}
Inclusive pion production spectra obtained from
Eichten {\it et~al.} \protect\cite{Eichten72} (solid circles),
Allaby {\it et~al.} \protect\cite{Allaby70} (triangles), and
Abbott {\it et al.} \protect\cite{Abbott92} (diamonds).
}
\end{figure}

In Fig.~\ref{xdndx-all} we show the inclusive pion
spectra obtained by integrating analytic parametrizations fitted to the 
data of Eichten {\it et~al.} \cite{Eichten72},
Allaby {\it et~al.} \cite{Allaby70} and Abbott {\it et~al.} \cite{Abbott92}.
The sensitivity of the results to different parametrizations used for
extrapolation is indicated by showing two points at the same $x$-value
for a given data
set. In addition to the typical systematic uncertainties of these 
experiments of
about 15\%, different methods of extrapolation result in pion spectra
which are different by up to 25\%. 

Finally, there are many proton-beryllium and proton-carbon particle
production measurements published which are restricted to one
or two angular bins of the secondaries (for example,
Baker {\it et~al.} \cite{Baker61}, 10, 20, 30 GeV; Dekkers {\it et~al.}
\cite{Dekkers65}, 19, 23 GeV). Due to the lack of 
theoretical understanding of soft hadronic particle production 
these data sets cannot be
used for inter- and extrapolation to obtain secondary pion spectra.


\section{Pion production other energies}

There are several measurements at lower energy, which cover
some relevant regions of phase
space~\cite{Angelov81,Baatar80a,Grigalashvili87,Agakishiyev85,Agakishiyev90,%
Armutliiski88,Ratner67,Bayukov79,Piroue66,Vorontsov83a,Vorontsov88a,Yamamoto81}.
In most of the cases only $\pi^-$ distributions have been measured and
some of the papers report on mean multiplicities or multiplicity
distributions only. Also, many data sets refer to thick targets and are
not directly usable for cascade calculations. In the following we
discuss only some of the published data sets where final state particles
have been identified and a large fraction of the phase space was
covered by the experiment. In particular, we do not consider the
numerous emulsion measurements.

At low energy, the data of
\cite{Agakishiyev85,Armutliiski88,Agakishiyev90} are interesting because
of the full coverage of the phase space.
\begin{figure}[!htb]
\centerline{\epsfig{figure=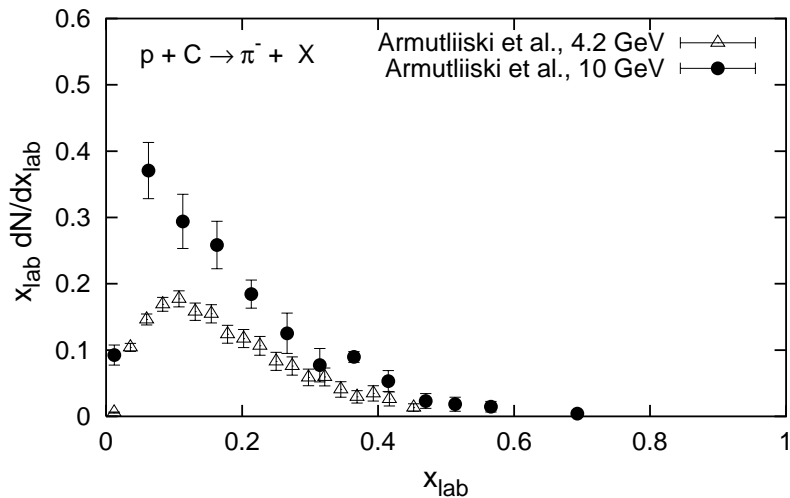,width=11cm}}
\caption{\label{dn-dxlab-pim-4}
Inclusive production spectrum of negative pions in proton-carbon
collisions \protect\cite{Agakishiyev85,Armutliiski88}.
}
\end{figure}
For example, in Fig.~\ref{dn-dxlab-pim-4} we show $\pi^-$
production spectra for proton-carbon collisions at 4.2 and 10 GeV. The
mean $\pi^-$ multiplicities are respectively $0.33\pm 0.02$ and
$1.00\pm0.03$ \cite{Agakishiyev85,Armutliiski88}\footnote{In
Ref.~\cite{Agakishiyev90} the 4.2 GeV data are also given as
double-differential distributions $Ed^3\sigma/dp^3$ in 10 degree bins of
the azimuthal angle.}
In addition, in Refs.~\cite{Agakishiyev85,Agakishiyev90} data on
$\alpha$-carbon and carbon-carbon collisions with 4.2 GeV/nucleon are
given.

The most relevant data sets at higher energy known to the authors are
in Refs.~\cite{Bozhko79a,Bozhko80b,Ambrosini99a}.
Ref.~\cite{Bozhko79a} gives the ratio of
particle production yields in proton-beryllium and proton-aluminum
collisions at 67 GeV. Such studies are important for the
extrapolation of p-Be data to p-air collisions. Whereas the measurements
published in \cite{Bozhko79a} refer to thin targets, the data of
Ref.~\cite{Bozhko80a} are for thick targets and both
thick and thin targets are considered in Ref.~\cite{Bozhko80b}.
The measurements of p-Be collisions at 450 GeV
of the NA56/SPY collaboration \cite{Ambrosini99a}
cover the important transverse
momentum range of $0 - 600$ MeV, however cross sections obtained by
extrapolation to a thin Be target are given only for the very forward
direction.


\section{Conclusion}

 After the new measurements of the primary cosmic ray spectrum,
 the uncertainties in the calculated sub-GeV neutrino fluxes
 are dominated by the limited information and understanding of
 hadronic interactions in the energy range from about 5 to 100 GeV.

 Most pressing is the need for a single experiment that covers
 pion production at several beam energies in the peak
 region of the sub-GeV events in Fig. 2, namely, around 20 GeV/c.
 Targets should be as close as possible to the constituents
 of the atmosphere, especially nitrogen.  If this is not possible,
 a series of targets with mass number spanning A=14 should be done.

At low energy, several data sets cover $\pi^-$ production. Data on
$\pi^+$ production is very sparse and many experiments refer to thick
targets or suffer from low statistics.
Thus,  a consistent set of new measurements extending to  
 lower energy would also be important, especially for
 interpreting measurements made at low geomagnetic cutoff.  

 Use of a beam of helium nuclei would also be of interest.
 In the energy range from 10 to 100 GeV/nucleon, approximately
 80\% of the primary cosmic-ray nucleons are free protons and 15\% 
 are bound in alpha
 particles.  Only 5\% or less are in heavier nuclei.


{\bf Acknowledgements.} This work is supported in part by U.S. Department
 of Energy contract DE~FG02~01ER~4062.





\end{document}